\documentclass[aps,pra,groupedaddress]{revtex4}

\usepackage{graphicx}
\begin{document}

%\tighten

\title{Bound entangled states violate a non-symmetric
local uncertainty relation}

\author{Holger F. Hofmann}
\email{h.hofmann@osa.org}
\author{}
%\altaffiliation{}
\affiliation{PRESTO, 
Japan Science and Technology Corporation 
(JST)\\
Research Institute for Electronic Science, Hokkaido 
University\\
Kita-12 Nishi-6, Kita-ku, Sapporo 060-0812, Japan}

\date{\today}

\begin{abstract}
As a consequence of having a positive partial transpose,
bound entangled states lack many of the properties otherwise
associated with entanglement. It is therefore interesting
to identify properties that distinguish bound entangled states 
from separable states. In this paper, it is shown that some 
bound entangled states violate a non-symmetric class of local 
uncertainty relations (quant-ph/0212090). This result indicates 
that the asymmetry of non-classical correlations may be a
characteristic feature of bound entanglement.
\end{abstract}

%\pacs{
%03.67.Mn %--Entanglement production, characterization and 
%         %  manipulation (see also 03.65.Ud)
%03.65.Ud %--Entanglement and quantum nonlocality 
%03.67.-a %--Quantum information
%\keywords{}

\maketitle

Entanglement is an essential element of quantum information
theory and is known to be responsible for a wide variety
of non-classical effects \cite{Bel64,Eke91,Ben92,Ben93,Ben96}. 
However, it is not generally clear what properties of
entanglement are required for any particular application. 
In fact, the properties of mixed 
entangled states are so difficult to characterize that even the 
question whether a given density matrix is entangled or 
separable may be difficult to answer 
\cite{Per96,Hor96,Hor97,Lew00,Guh02}. 
For $2\times 2$ and $2\times 3$
systems, this problem can be solved by testing whether the
partial transpose of the density matrix is positive or not
\cite{Per96,Hor96}. However, for higher dimensional Hilbert spaces,
there are examples of entangled states with a positive partial 
transpose \cite{Hor97,Hor98,Hor00}. 
Since the positivity of the partial transpose
is a fundamental non-local property of the state that cannot
be changed using only local operations and classical 
communication, the entanglement of states with positive partial
transpose cannot be distilled to singlett form and is therefore
not available for standard applications in quantum information
protocols \cite{Hor98}. For this reason, entangled states with
positive partial transpose are generally referred to as bound 
entangled states. Nevertheless, the inseparability of bound 
entanglement
suggests that there are other fundamental properties of
entanglement that distinguish bound entanglement from separable
states. In particular, it would be interesting to know whether
there are some potentially useful properties of bound entanglement
that can be applied without requiring distillation to pure states.
In this paper, it is shown that some bound entangled states 
do indeed have such a property. Specifically, the correlations
between the two systems in the $3 \times 3$ bound entanglement
analyzed in the following overcome the local uncertainty limit
\cite{Hof02}. This means that bound entanglement could be 
applied directly in quantum communication, e.g. for 
quantum teleportation or for dense coding, since the amount 
of noise in data transmission 
would clearly be less than that of the corresponding 
classical limit.

The type of bound entanglement considered in this paper
is the $3\times 3$ entanlgement first presented in \cite{Hor97}.
Since the angular momentum components of the spin-1 algebra
will be used in the analysis, it is most convenient to express
this state in the $\hat{l}_z$ basis as
\begin{eqnarray}
\label{eq:bound}
\hat{\rho}_a&=& \frac{a}{1+ 8 a} \big( 
   \mid -1; 0 \rangle\langle -1; 0 \mid +
   \mid -1;+1 \rangle\langle -1;+1 \mid +
   \mid  0;-1 \rangle\langle  0;-1 \mid +
   \mid  0;+1 \rangle\langle  0;+1 \mid +
   \mid +1; 0 \rangle\langle +1; 0 \mid \big)
\nonumber \\ && + \frac{3 a}{1+ 8 a} 
   \mid E_{\mbox{max.}} \rangle\langle E_{\mbox{max.}} \mid
                + \frac{1}{1+ 8 a} 
   \mid \Pi \rangle\langle \Pi \mid,
\nonumber \\[0.3cm] && \mbox{where} \hspace{0.5cm} 
\mid E_{\mbox{max.}} \rangle = 
\frac{1}{\sqrt{3}}\left(
\mid -1;-1 \rangle + \mid 0;0 \rangle + \mid +1;+1\rangle
\right)
\nonumber \\ && \hspace{0.2cm} 
\mbox{and} \hspace{0.7cm} \mid \Pi \rangle =
\sqrt{\frac{1+a}{2}} \mid +1;-1 \rangle + 
\sqrt{\frac{1-a}{2}} \mid +1;+1 \rangle.
\end{eqnarray}
The parpameter $a$ can take any value between zero and one.
In order to analyze the correlations between the physical 
properties of the two three level systems represented by this 
density matrix, it is necessary to express the statistics
of the density matrix in terms of expectation values of 
observables. A particularily convenient description of this 
type can be obtained by using a set of eight Hermitian 
generating operators $\hat{\lambda}_i$ characterized by the 
relations \cite{Mahler}
\begin{eqnarray}
\label{eq:generate}
\mbox{Tr}\{\hat{\lambda}_i \} &=& 0
\nonumber \\
\mbox{Tr}\{\hat{\lambda}_i\hat{\lambda}_j \} &=& 2 \delta_{i,j}
\\
\label{eq:square}
\sum_i \hat{\lambda}_i^2 &=& \frac{16}{3}\; \hat{1}.
\end{eqnarray}
The expectation values of these generating operators can then 
be interpreted as a generalization of the Bloch vector. 
In particular, the purity of the density matrix of a three
level system can be expressed as
\begin{equation}
\label{eq:purity}
\mbox{Tr}\{\hat{\rho}^2_{\mbox{local}} \} =
\frac{1}{3} + 
\frac{1}{2} \sum_i \langle \hat{\lambda}_i \rangle^2.
\end{equation}
This relation implies that the length of the eight dimensional
Bloch vector is limited to $\sqrt{4/3}$. Using equations
(\ref{eq:square}) and (\ref{eq:purity}), it is then possible to 
formulate the sum uncertainty relation \cite{Hof02} for the 
generating operators $\hat{\lambda}_i$, 
\begin{equation}
\label{eq:purlimit}
\sum_i \delta \lambda_i^2 \geq 4.
\end{equation}
As explained in \cite{Hof02}, this purity uncertainty can be 
used to define a sufficient condition for entanglement. 
Specifically, no separable state of the $3 \times 3$ system
can violate any local uncertainty relation of the form
\begin{equation}
\label{eq:plur}
\sum_i \delta\left(\lambda_i(1) - \lambda_i(2)\right)^2 \geq 8,
\end{equation}
where $\lambda_i(1/2)$ represent the measurement outcomes for
the observables corresponding to the operators 
$\hat{\lambda}_i(1/2)$, respectively. It should be noted, however,
that $\hat{\lambda}_i(1)$ and $\hat{\lambda}_i(2)$ do not have to
be the same operators. Indeed, it is an important part of the 
result presented in this paper that they can have completely
different properties. 

It is now possible to define an optimal selection of operators
$\hat{\lambda}_i(1)$ and $\hat{\lambda}_i(2)$ for the bound 
entangled state (\ref{eq:bound}). In order to obtain both
a compact formulation and a direct connection with the physical
properties of a spin 1 system, one fundamental set of generating 
operators can be defined using the operators of the spin 
components, $\hat{l}_x$, $\hat{l}_y$, and $\hat{l}_z$, and their
quadratic functions,
\begin{eqnarray}
\label{eq:definitions}
\hat{Q}_{ij} &=& \hat{l}_i\hat{l}_j+\hat{l}_j\hat{l}_i
\nonumber \\
\hat{S}_{xy} &=& \hat{l}_x^2 - \hat{l}_y^2
\nonumber \\
\hat{G}_{z} &=& \sqrt{3} \left(\hat{l}_z^2-\frac{2}{3}\right).
\end{eqnarray}
With these basic definitions, the correlations of the bound
entanglement described by (\ref{eq:bound}) can be expressed
in terms of optimally aligned operator pairs. This optimal
alignment is determined by maximizing the total correlation
given by
\begin{equation}
\label{eq:K}
K_{\mbox{total}}=\sum_i \langle \hat{\lambda}_i(1)
\otimes \hat{\lambda}_i(2) \rangle.
\end{equation}
The choices of $\hat{\lambda}_1(1)$ to $\hat{\lambda}_5(1)$
are determined by the fact that these operator properties 
are only correlated with the respective operator properties
$\hat{\lambda}_1(2)$ to $\hat{\lambda}_5(2)$ in system 2.
However, there are some cross correlations in the 
remaining three operator properties, making it necessary to 
determine a non-trivial selection of operators.
In general, the optimized operator alignment can be 
given by the following set of operator pairs,
\begin{eqnarray}
\label{eq:lambdas}
\hat{\lambda}_1(1) = \hat{l}_x(1) && 
\hat{\lambda}_1(2) = \hat{l}_x(2)
\nonumber \\ 
\hat{\lambda}_2(1) = -\hat{l}_y(1) && 
\hat{\lambda}_2(2) =  \hat{l}_y(2)
\nonumber \\ 
\hat{\lambda}_3(1) = -\hat{Q}_{xy}(1) && 
\hat{\lambda}_3(2) =  \hat{Q}_{xy}(2)
\nonumber \\ 
\hat{\lambda}_4(1) = -\hat{Q}_{yz}(1) && 
\hat{\lambda}_4(2) =  \hat{Q}_{yz}(2)
\nonumber \\ 
\hat{\lambda}_5(1) = \hat{Q}_{zx}(1) && 
\hat{\lambda}_5(2) = \hat{Q}_{zx}(2)
\nonumber \\ 
\hat{\lambda}_6(1) = \hat{Z}(1) && 
\hat{\lambda}_6(2) = \hat{l}_z(2)
\nonumber \\ 
\hat{\lambda}_7(1) = \hat{F}_{xy}(1) && 
\hat{\lambda}_7(2) = \hat{S}_{xy}(2)
\nonumber \\ 
\hat{\lambda}_8(1) = \hat{F}_z(1) && 
\hat{\lambda}_8(2) = \hat{G}_z(2),
\end{eqnarray}
where $\hat{Z}(1)$, $\hat{F}_{xy}(1)$, and $\hat{F}_z(1)$
are linear combinations of $\hat{l}_z(1)$, $\hat{S}_{xy}(1)$,
and $\hat{G}_z(1)$ that have to be optimized depending on the
specific value of $a$ chosen for $\hat{\rho}$ in 
(\ref{eq:bound}). The result of this optimization reads
\begin{eqnarray}
\label{eq:asym}
\hat{Z} &=& \frac{1}{2} \hat{l}_z + \frac{\sqrt{3}}{2} \hat{G}_z
\nonumber \\
\hat{F}_{xy} &=& 
\frac{1+2 a}{2+a} \hat{S}_{xy} +
\frac{\sqrt{3 (1-a^2)}}{2+a}
\left(\frac{\sqrt{3}}{2} \hat{l}_z - \frac{1}{2} \hat{G}_z\right)
\nonumber \\
\hat{F}_{z} &=& 
\frac{1+2 a}{2+a} \left(
\frac{\sqrt{3}}{2} \hat{l}_z - \frac{1}{2} \hat{G}_z\right) 
- \frac{\sqrt{3 (1-a^2)}}{2+a} \hat{S}_{xy}.
\end{eqnarray}
With this choice of operator properties, the maximal
correlation achieved is always equal to $K_{\mbox{total}}=4/3$.
Since this result is exactly equal to the square of the 
maximal length of the local Bloch vector, the optimized 
correlation already corresponds to the maximal correlation 
that can be achieved in separable systems. 
The local uncertainty defined by (\ref{eq:plur}) and
(\ref{eq:lambdas}) can now be evaluated by using this 
correlation,
\begin{eqnarray}
\label{eq:vio}
\sum_i \delta\left(\lambda_i(1) - \lambda_i(2)\right)^2 &=&
\underbrace{
\left(\sum_i 
\langle (\hat{\lambda}_i(1) - \hat{\lambda}_i(2))^2\rangle
\right)}_{32/3 - 2 K_{\mbox{\tiny total}}} 
- \left(\sum_i 
\langle \hat{\lambda}_i(1) - \hat{\lambda}_i(2)\rangle^2
\right) 
\nonumber \\
&=& 8 - \left(\sum_i 
\langle \hat{\lambda}_i(1) - \hat{\lambda}_i(2)\rangle^2
\right) < 8.
\end{eqnarray}
The local uncertainty relation (\ref{eq:plur}) is therefore
violated because of the non-vanishing mismatch in the local 
expectation values given by
\begin{eqnarray}
\label{eq:mis}
\langle \lambda_7(1)-\lambda_7(2) \rangle
&=& - \frac{3 a \sqrt{1-a^2}}{(2+a)(1+8a)}
\nonumber \\
\langle \lambda_8(1)-\lambda_8(2) \rangle
&=& \frac{\sqrt{3} a (1-a)}{(2+a)(1+8a)}.
\end{eqnarray}
Any separable states with a maximal correlation total
of $K_{\mbox{total}}=4/3$ must have perfectly aligned local
Bloch vectors. However, the mismatch given by equation
(\ref{eq:mis}) shows that this is not the case for the
bound entangled state $\rho_a$ given by equation 
(\ref{eq:bound}). Therefore, this class of bound entangled 
states violates the local uncertainty relation 
(\ref{eq:plur}).

\begin{figure}
\begin{picture}(300,170)
%\put(0,0){\framebox(300,170){}}
\put(30,15){\makebox(250,150){\includegraphics{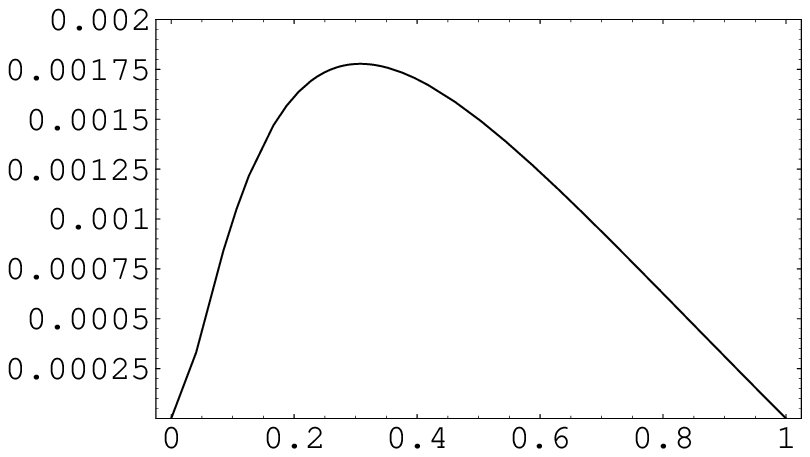}}}
\put(0,85){\makebox(40,20){\large $C_{\mbox{\normalsize LUR}}$}}
\put(168,2){\makebox(20,20){\large $a$}}
\end{picture}
\caption{\label{clur} Relative violation of local uncertainty
$C_{\mbox{\small LUR}}$ as a function of the parameter $a$ 
defining the bound entangled state $\hat{\rho}_a$.}
\end{figure}
As discussed in \cite{Hof02}, a useful measure of the relative 
violation of local uncertainty can be obtained by normalizing
the amount by which the uncertainty is violated with the
uncertainty limit,
\begin{eqnarray}
\label{eq:clur}
C_{\mbox{\small LUR}} &=& 1-\frac{1}{8}
\sum_i \delta\left(\lambda_i(1) - \lambda_i(2)\right)^2
\nonumber \\
&=& \frac{1}{8}\left(\langle \lambda_7(1)-\lambda_7(2) \rangle^2
+ \langle \lambda_8(1)-\lambda_8(2) \rangle^2\right)
\nonumber \\
&=& \frac{3 a^2(1-a)}{4 (2+a)(1+8a)^2}.
\end{eqnarray}
The dependence of this relative violation of local uncertainty
on the parameter $a$ that defines the bound entangled state is
shown in figure \ref{clur}. This result shows that the 
non-classical correlations of the bound entangled states
considered here are about one thousand times weaker than the
non-classical correlations of maximally entangled states.
It is also interesting to note that the maximal amount of
bound entanglement is obtained for $a \approx 0.3077$, with
a relative local uncertainty violation of 
$C_{\mbox{\small LUR}} \approx 0.00178$.  

Using the violation of local uncertainty as a criterion, it 
is also possible to extend the class of bound entangled 
states given by equation (\ref{eq:bound})
to mixtures of $\hat{\rho}_a$ and white noise,
\begin{equation}
\hat{\rho}(a;p_N) = p_N \hat{1} \otimes \hat{1}
+ (1-p_N) \hat{\rho}_a.
\end{equation}
Such states still violate the local uncertainty relation
(\ref{eq:plur}) as long as the noise level is below the
limit given by
\begin{equation}
\frac{p_N}{3(1-p_N)^2} < C_{\mbox{\small LUR}}(p_N=0),
\end{equation}
where $C_{\mbox{\small LUR}}(p_N=0)$ is the relative violation
of local uncertainty for $\hat{\rho}_a$ given by equation
(\ref{eq:clur}) above. This result indicates that an addition
of noise to bound entanglement is not critical if the noise level
is well below $0.5 \%$, thus determining the level of 
precision required for an experimental investigation of bound 
entangled states. 

Besides the possibility of quantifying the non-classical 
properties of bound entanglement, the violation of local 
uncertainty relations also provides insights into the 
physical properties of bound entanglement. The local uncertainty
relation (\ref{eq:plur}) defines a correspondence between the
properties $\hat{\lambda}_i(1)$ of system one and the properties
$\hat{\lambda}_i(2)$ of system two. However, equation 
(\ref{eq:asym}) defines the last three operator pairs in a
highly asymmetric fashion. In particular, the correlated 
operators do not even have the same eigenvalue spectrum. 
It may well be that this lack of symmetry in bound entanglement
is the main practical obstacle preventing the construction of
entanglement purification protocols for bound entangled states
\cite{Hor98}. Nevertheless the fact that bound entanglement 
overcomes the uncertainty limit given by relation 
(\ref{eq:plur}) suggests that it may actually be used 
directly to realize a kind of quantum teleportation. 
Specifically, quantum teleportation using bound entanglement 
would transfer the properties of the input state to 
properties of the output state according to a trace 
preserving map defined by the Bell measurement and the pair 
correlations given by (\ref{eq:lambdas}). Since this map 
changes the eigenvalue spectrum of operators, it is 
necessarily non-positive. The positivity of the output state 
is only preserved by the noise added in the transfer process. 
Even though this kind of asymmetric teleportation therefore
requires a certain minimum of noise, the violation of the local 
uncertainty relation (\ref{eq:plur}) shows that the transfer of
properties would still be more precise than local operations and
classical communication would allow. Quantum teleportation 
using bound entangled states can thus be seen as a non-classical
implementation of a non-positive map, similar to quantum
cloning or the universal NOT operation \cite{Buz96,Gis97,Buz99}.

In conclusion, it has been shown that the bound entangled states
given by (\ref{eq:bound}) violate the local uncertainty relation
(\ref{eq:plur}) defined by the choice of generating operators
(\ref{eq:lambdas}). It is then possible to quantify the amount
of entanglement in terms of the relative violation of local
uncertainty and to identify the non-classical correlations
between the two systems in terms of local physical properties. 
The result indicates a specific kind of asymmetry between 
the correlated operator properties, characterized by the fact
that the correlated operators do not share the same eigenvalue 
spectrum. It may well be that this kind of asymmetry is largely 
responsible for the lack of distillability in bound 
entanglement. However, the violation of local uncertainty itself
shows that bound entanglement may overcome the classical limit
in applications such as quantum teleportation or dense coding.

\end{document}